\documentclass[pra,byrevtex,superscriptaddress,twocolumn]{revtex4}%
\usepackage{amsfonts}
\usepackage{amsmath}
\usepackage{amssymb}
\usepackage{graphicx}%
\begin{document}
\preprint{ }
\title[Short title for running header]{Theory of Dicke narrowing in coherent population trapping}
\author{O. Firstenberg}
\affiliation{Department of Physics, Technion-Israel Institute of Technology, Haifa 32000, Israel}
\author{M. Shuker}
\affiliation{Department of Physics, Technion-Israel Institute of Technology, Haifa 32000, Israel}
\author{A. Ben-Kish}
\affiliation{Department of Physics, Technion-Israel Institute of Technology, Haifa 32000, Israel}
\author{D. R. Fredkin}
\affiliation{Department of Physics, University of California, San Diego, La Jolla,
California 92093}
\author{N. Davidson}
\affiliation{Department of Physics of Complex Systems, Weizmann Institute of Science,
Rehovot 76100, Israel }
\author{A. Ron}
\affiliation{Department of Physics, Technion-Israel Institute of
Technology, Haifa 32000, Israel}


\begin{abstract}
The Doppler effect is one of the dominant broadening mechanisms in
thermal vapor spectroscopy. For two-photon transitions one would
naively expect the Doppler effect to cause a residual broadening,
proportional to the wave-vector difference. In coherent population
trapping (CPT), which is a two-photon narrow-band phenomenon, such
broadening was not observed experimentally. This has been commonly
attributed to frequent velocity-changing collisions, known to
narrow Doppler-broadened one-photon absorption lines (Dicke
narrowing). Here we show theoretically that such a narrowing
mechanism indeed exists for CPT resonances. The narrowing factor
is the ratio between the atom's mean free path and the wavelength
associated with the wave-vector difference of the two radiation
fields. A possible experiment to verify the theory is suggested.

\end{abstract}
\maketitle

\section{Introduction}

\begin{figure}
[ptb]
\begin{center}
\includegraphics[
height=1.4036in,
width=3.1981in
]%
{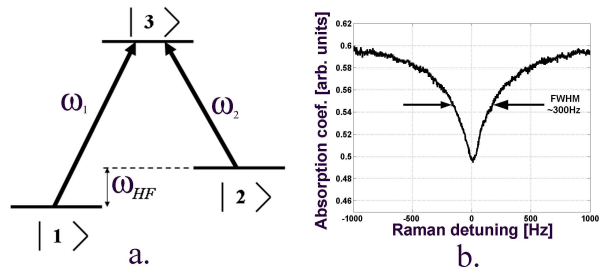}%
\caption{a. Energy levels scheme of CPT system. Two laser fields with
frequencies $\omega_{1},\omega_{2}$ excite two levels to a common upper level.
b. Typical measurement of a CPT resonance\ in room temperature $^{87}$Rb
vapor, with frequency difference of $\omega_{\text{HF}}=6.834$GHz. The
residual Doppler broadening is about $10$KHz, while the measured width is only
about $300$Hz.}%
\label{figure1}%
\end{center}
\end{figure}

The spectral line shape of atomic transitions is determined by
many different mechanisms \cite{Dicke1956}. Specifically, a
Doppler-broadened spectrum can be dramatically narrowed due to
frequent velocity-changing collisions --- Dicke narrowing
\cite{Dicke1953,GalatryPR1961}. The narrowing factor is
proportional to the ratio between the collisions mean free path
and the radiation wavelength. Dicke narrowing was observed for
microwave transitions \cite{budker2005} and recently also for
optical transitions \cite{Dutier2003}. For two-photon transitions,
such as coherent population trapping (CPT) \cite{Arimondo96}, a
dramatic narrowing of the expected Doppler-width was also observed
and was attributed to a Dicke-like narrowing
\cite{Cyr1993,WynandsPRA1999,VanierPRA2003,dutier2005}. CPT is a
light-matter interaction involving three-level atoms and two
resonant radiation fields (see figure \ref{figure1}.a). The atoms
are driven into a coherent superposition of the lower levels,
$\left\vert 1\right\rangle$ and $\left\vert 2\right\rangle$ which
does not absorb the radiation. The efficiency of this process
strongly depends on the Raman (two-photon) detuning, defined as
$\Delta_{R}=\omega_{\text{HF}}-(\omega_{1}-\omega_{2})$.
Therefore, when scanning the frequency of one of the lasers, an
absorption dip is observed, which we will refer to as a CPT
resonance (see figure \ref{figure1}.b).

The narrow width of CPT resonances is useful for several
applications such as frequency standards
\cite{Cyr1993,knappe:1460}, magnetometers \cite{schwindt:6409},
slow and stored light \cite{LukinRMP2003,FleischhauerRMP2005}. In
all the applications a key parameter is the spectral width of the
CPT resonance. When the CPT is performed in room-temperature
vapor, the two radiation fields experience a Doppler shift which
is different for each atomic velocity group. In the case of
non-degenerate lower levels (e.g. two hyperfine levels in the
ground state of an alkali atom) or for non-collinear laser beams,
the two radiation fields experience slightly different Doppler
shifts. This difference, denoted as the \emph{residual} Doppler
shift, results in an effective Raman detuning which is different
for each velocity group. Therefore, one would naively expect a
residual Doppler broadening of the CPT resonance. However, the
measured CPT resonance width is well below the expected residual
Doppler width. For example, the clock-transition CPT\ resonance of
room temperature Rubidium
vapor is expected to have a residual Doppler width of%
\begin{equation}
\Gamma_{D}^{\text{res}}=\frac{v_{\text{th}}}{c}\omega_{\text{HF}}%
\approx10\text{KHz,}%
\end{equation}
where $v_{\text{th}}$ is the thermal velocity,
$\omega_{\text{HF}}$ is the hyperfine energy-gap, and $c$ is the
speed of light. However, a typical measurement of this CPT
resonance, depicted in figure \ref{figure1}.b, shows a width of
only $300$Hz. In order to attribute this dramatic narrowing to a
Dicke-like narrowing effect, one commonly assumes that the
wavelength which determines the narrowing factor is the one
associated with the \emph{frequency difference} of the lasers
(i.e. the microwave frequency) \cite{Cyr1993}. In Ref.
\cite{HelmPRA2001} a numerical model was used to calculate the CPT
line-shape, introducing discrete velocity groups, and indeed
demonstrated the expected narrowing. In the present work we derive
an analytic expression for the line-shape of CPT resonances, that
demonstrates both Doppler broadening and collisions-induced Dicke
narrowing. We show that in a regime relevant to most CPT
experiments, the narrowing is governed by the ratio between the
mean free path and the wavelength associated with the wave-vector
difference of the lasers. In section II we review the two-level
Dicke narrowing. In section III we derive the theory for Dicke
narrowing of CPT resonances. Section IV contains a discussion of
the results and presents a possible experimental setup to verify
this result.

\section{Dicke narrowing of two-level atoms absorption line}

In this section, as an introduction to the CPT case, we develop a theory for
the line-shape of a two-level atom, including both Doppler broadening and
Dicke narrowing. Our approach is similar to that presented by Galatry
\cite{GalatryPR1961}. We consider a two-level atom with the upper and lower
states, $\left\vert 3\right\rangle $ and $\left\vert 1\right\rangle $, and an
optical energy gap of $\hbar\omega_{0}$. We assume that the kinetic energy of
the atom is much smaller than $\hbar\omega_{0}$ and take the motion of the
center of mass of the atom to be classical. The atom interacts with an
external, classical electromagnetic field, with wave-vector $\mathbf{q}$ and
frequency $\omega=c\left\vert \mathbf{q}\right\vert $. We assume that $\omega$
is of the order of $\omega_{0}$ and express the Hamiltonian in the dipole
approximation and the rotating wave approximation as%

\begin{equation}
H=\hbar\omega_{0}\left\vert 3\right\rangle \left\langle 3\right\vert -\left[
\hbar\Omega e^{i\left(  \mathbf{q\cdot r}-\omega t\right)  }\left\vert
3\right\rangle \left\langle 1\right\vert +\text{H.c.}\right]  \text{,}%
\end{equation}
where $\Omega$ is the Rabi frequency of the field and $\mathbf{r}%
=\mathbf{r}\left(  t\right)  $ is the time-dependent center of mass position
of the atom. The equations of motion of the density matrix elements
$\rho\,_{3,3}$ and $\rho_{3,1}$ are given by:%
\begin{subequations}
\begin{align}
\dot{\rho}_{3,1}  &  =-i\left(  \omega_{0}-i\Gamma\right)  \rho_{3,1}-i\Omega
e^{i\left(  \mathbf{q\cdot r}-\omega t\right)  }\left(  2\rho_{3,3}-1\right)
\\
\dot{\rho}_{3,3}  &  =-2\Gamma\rho_{3,3}+2\operatorname{Im}\Omega^{\ast
}e^{-i\left(  \mathbf{q\cdot r}-\omega t\right)  }\rho_{3,1}.
\label{2l_sigma_z3}%
\end{align}
Here we assumed a single relaxation term (radiation bath), which induces
transitions between the atomic levels $\left(  \left\vert 3\right\rangle
\rightarrow\left\vert 1\right\rangle \right)  ,$ with the rate $\Gamma
\ $\cite{AtomPhotonInteractions}. In order to calculate the absorption
spectrum, we write the energy absorption $\dot{W}\left(  t\right)  ,$ in terms
of the population response to the applied field:%

\end{subequations}
\begin{equation}
\dot{W}\left(  t\right)  =\hbar\omega_{0}\left\{  \dot{\rho}_{3,3}\right\}
_{\text{field}}=2\hbar\omega_{0}\operatorname{Im}\left(  \Omega^{\ast
}e^{-i\left(  \mathbf{q\cdot r}-\omega t\right)  }\rho_{3,1}\right)  .
\label{2l_wdot}%
\end{equation}
In the steady-state we take the temporal average, and find the absorption
spectrum $S\left(  \omega\right)  $ to be:%
\begin{equation}
S\left(  \omega\right)  =\frac{\overline{\dot{W}\left(  t\right)  }}%
{2\hbar\omega_{0}\left\vert \Omega\right\vert ^{2}}=\operatorname{Im}%
\lim_{T\rightarrow\infty}\int_{0}^{T}\frac{dt}{T}\frac{\rho_{3,1}\left(
t\right)  }{\Omega e^{i\left(  \mathbf{q\cdot r}-\omega t\right)  }}.
\label{2l_s}%
\end{equation}

We consider the unsaturated case in which $\Omega\ll\Gamma$, and use the
perturbation expansion of $\dot{\rho}\left(  t\right)  $. Taking the initial
state of the atom to be the ground state, we find to zero order in $\Omega$
that $\rho_{3,3}^{(0)}\left(  t\right)  =\rho_{3,1}^{(0)}\left(  t\right)
=0$, and to the first order that
\begin{equation}
\dot{\rho}_{3,1}^{\left(  1\right)  }=-i\left(  \omega_{0}-i\Gamma\right)
\rho_{3,1}^{\left(  1\right)  }-i\Omega e^{i\left(  \mathbf{q\cdot r}-\omega
t\right)  }\left(  2\rho_{3,3}^{(0)}-1\right)  ,\label{2l_firstorder}%
\end{equation}
with the formal solution%
\begin{subequations}
\begin{equation}
\rho_{3,1}^{\left(  1\right)  }\left(  t\right)  =i\Omega\int_{0}%
^{t}dt^{\prime}e^{-i\left(  \omega_{0}-i\Gamma\right)  \left(  t-t^{\prime
}\right)  }e^{i\mathbf{q\cdot r}\left(  t^{\prime}\right)  -i\omega t^{\prime
}}.\label{2l_rho31_1}%
\end{equation}
Substituting $\rho_{3,1}^{\left(  1\right)  }\left(  t\right)  $ in
Eq.(\ref{2l_s}), denoting $\tau=t-t^{\prime}$ and taking the upper limit of
the $\tau-$integral to infinity, we obtain in steady-state,
\end{subequations}
\begin{equation}
S\left(  \omega\right)  =\operatorname{Re}\int_{0}^{\infty}d\tau e^{-i\left(
\Delta-i\Gamma\right)  \tau}\overline{e^{i\Phi\left(  \tau\right)  }%
},\label{2l_s3}%
\end{equation}
where $\Delta=\omega-\omega_{0}$ is the detuning from resonance,%
\begin{equation}
\Phi\left(  \tau\right)  =\mathbf{q\cdot}\left[  \mathbf{r}\left(
\tau\right)  -\mathbf{r}\left(  0\right)  \right]  ,\label{2l_Phi_t}%
\end{equation}
is the phase accumulated during $\tau$ and the average is over atom
trajectories,
\begin{equation}
\overline{e^{i\Phi\left(  \tau\right)  }}=\lim_{T\rightarrow\infty}\frac{1}%
{T}\int_{0}^{T}dte^{i\mathbf{q\cdot}\left[  \mathbf{r}\left(  t\right)
-\mathbf{r}\left(  t-\tau\right)  \right]  }.
\end{equation}
Notice that the accumulated phase includes both the Doppler effect and
collisions. Assuming that $\Phi\left(  \tau\right)  $ is a random Gaussian
variable and following the cumulant expansion procedure \cite{AS928}, we
substitute $\overline{e^{i\Phi\left(  \tau\right)  }}=e^{-\overline
{\Phi\left(  \tau\right)  ^{2}}/2}.$ Following Eq.(\ref{2l_Phi_t}) we write%
\begin{equation}
\Phi\left(  \tau\right)  =\sum_{\alpha}q^{\alpha}\int_{o}^{\tau}%
dt_{1}u^{\alpha}\left(  t_{1}\right)  ,\label{2l_single_phi}%
\end{equation}
where $u^{\alpha}$ are the velocity cartesian components and $\alpha=x,y,z$,
to get
\begin{align}
\overline{\Phi\left(  \tau\right)  ^{2}} &  =\sum_{\alpha,\alpha^{\prime}%
}q^{\alpha}q^{\alpha^{\prime}}\int_{0}^{\tau}dt_{1}\int_{0}^{\tau}%
dt_{2}\overline{u^{\alpha}\left(  t_{1}\right)  u^{\alpha^{\prime}}\left(
t_{2}\right)  }\nonumber\\
&  =2\sum_{\alpha,\alpha^{\prime}}q^{\alpha}q^{\alpha^{\prime}}\int_{0}^{\tau
}dt\left(  \tau-t\right)  \overline{u^{\alpha}\left(  t\right)  u^{\alpha
^{\prime}}\left(  0\right)  }.\label{2l_phisqr0}%
\end{align}

As shown in the appendix, the velocity-velocity correlation function is given
approximately by%
\begin{equation}
\overline{u^{\alpha}\left(  t\right)  u^{\alpha^{\prime}}\left(  0\right)
}=\delta_{\alpha^{\prime}\alpha}v_{\text{th}}^{2}e^{-\gamma\left\vert
t\right\vert },\label{2l_uu_corr}%
\end{equation}
where $\delta_{\alpha,\alpha^{\prime}}$ is the Kronecker Delta, $v_{\text{th}%
}$ is the thermal velocity and $\gamma$ is interpreted as the velocity
relaxation rate (for the Brownian motion regime) or as the\ collisions rate
(for the strong collisions regime). Substituting this in Eq.(\ref{2l_phisqr0})
gives%
\begin{equation}
\overline{\Phi\left(  \tau\right)  ^{2}}=2q^{2}v_{\text{th}}^{2}\int_{0}%
^{\tau}dt\left(  \tau-t\right)  e^{-\gamma t}=\frac{2q^{2}v_{\text{th}}^{2}%
}{\gamma^{2}}G\left(  \gamma\tau\right)  \label{2l_phisqr}%
\end{equation}
where
\begin{equation}
G\left(  x\right)  =x-1+e^{-x}.\label{2l_Gx}%
\end{equation}
Then the absorption spectrum, Eq.(\ref{2l_s3}), is written as
\begin{equation}
S\left(  \omega\right)  =\int_{0}^{\infty}d\tau e^{-\Gamma\tau-\Gamma_{D}%
^{2}G\left(  \gamma\tau\right)  /\gamma^{2}}\cos\left(  \Delta\tau\right)
,\label{2l_s_final}%
\end{equation}
where $\Gamma_{D}=qv_{\text{th}}=\omega v_{\text{th}}/c$ is the Doppler width.
Restricting the discussion to the regime where $\Gamma_{D}\gg\Gamma,$ one
finds two limits \cite{GalatryPR1961}:

(i) The Doppler limit, $\gamma\ll\Gamma_{D}$, where the quadratic term
$G\left(  \gamma\tau\right)  \approx\gamma^{2}\tau^{2}/2$ is dominant and the
absorption spectrum,%
\begin{equation}
S_{\text{Doppler}}\left(  \omega\right)  =\sqrt{\frac{\pi}{2}}\frac{1}%
{\Gamma_{D}}e^{-\frac{\Delta^{2}}{2\Gamma_{D}^{2}}},
\end{equation}
is the well-known Doppler-broadened spectrum. The same expression is usually
derived by a convolution of the homogenous line shape with the thermal
velocity distribution.

\begin{figure}
[ptb]
\begin{center}
\includegraphics[
height=2.5503in,
width=3.3918in
]%
{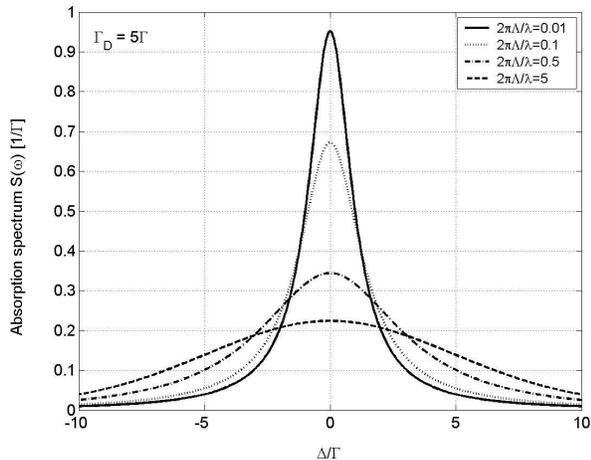}%
\caption{The effect of Dicke narrowing on the absorption spectrum: a numerical
solution of Eq.(\ref{2l_s_final}) for several values of the Dicke parameter
$2\pi\Lambda/\lambda$, with $\Gamma_{D}=5\Gamma.$}%
\label{2l_s_shapes}%
\end{center}
\end{figure}

(ii) The Dicke limit, $\gamma\gg\Gamma_{D}$, where the linear term $G\left(
\gamma\tau\right)  \approx\gamma\tau$ is dominant and the absorption spectrum,%
\begin{equation}
S_{\text{Dicke}}\left(  \omega\right)  =\frac{\Gamma+\Gamma_{D}^{2}/\gamma
}{\Delta^{2}+\left(  \Gamma+\Gamma_{D}^{2}/\gamma\right)  ^{2}}%
,\label{2l_sdicke}%
\end{equation}
is a Lorentzian with natural-width $\Gamma$, broadened by $\Gamma_{D}%
^{2}/\gamma.$ Denoting the mean free path $\Lambda=v_{\text{th}}/\gamma,$ and
the field's wavelength $\lambda=2\pi/q$, we find the line-width to be
$\Gamma+\left(  2\pi\Lambda/\lambda\right)  \Gamma_{D},$ showing that the
Doppler broadening is effectively reduced by a factor known as the Dicke
parameter \cite{Dicke1953}. A numerical solution of Eq.(\ref{2l_s_final}),
showing the transition between the Doppler and the Dicke limits, is shown in
Fig.(\ref{2l_s_shapes}).%

\section{Dicke narrowing of CPT resonances}

We now turn to analyze the line-shape of CPT resonances in the presence of
thermal motion and velocity-changing collisions. Consider a three-level atom
in a $\Lambda-$configuration, with the energy levels $E_{3}>E_{2}>E_{1}$
corresponding to an upper state $\left\vert 3\right\rangle $ and two lower
states $\left\vert 2\right\rangle ,\left\vert 1\right\rangle $ (see figure
\ref{figure1}.a). We will denote the atomic frequencies $\omega_{nn^{\prime}%
}=\left(  E_{n}-E_{n^{\prime}}\right)  /\hbar.$ The atom interacts with
external probe and pump electric fields, close to resonance with the
$\left\vert 3\right\rangle \rightarrow\left\vert 1\right\rangle $ and
$\left\vert 3\right\rangle \rightarrow\left\vert 2\right\rangle $ transitions,
respectively. The coupling Hamiltonian is%
\begin{equation}
H_{C}=-\hbar\sum_{n=1,2}\Omega_{n}e^{i\mathbf{q}_{n}\mathbf{\cdot r}\left(
t\right)  -i\omega_{n}t}\left\vert 3\right\rangle \left\langle n\right\vert
+\text{H.c.}%
\end{equation}
where $\mathbf{q}_{n}$ are the wave-vectors of the fields and $\omega
_{n}=c\left\vert \mathbf{q}_{n}\right\vert $. For brevity, we denote%
\begin{equation}
\Omega_{n}\left(  t\right)  =\Omega_{n}e^{i\mathbf{q}_{n}\mathbf{\cdot
r}\left(  t\right)  -i\omega_{n}t}~~~~n=1,2.\label{Omega_n}%
\end{equation}
We consider both \emph{dipole} relaxation terms, which induce transitions
between the excited and ground levels$~\left\vert 3\right\rangle
\rightarrow\left\vert n\right\rangle $ with rates $\Gamma_{n}$ ($n=1,2$), and
within the ground state $\left\vert 1\right\rangle \leftrightarrow\left\vert
2\right\rangle $ with rate $\Gamma_{2\leftrightarrow1}$; and a
\emph{spin-boson} relaxation term, which induces adiabatic transitions
(decoherence) between $\left\vert 1\right\rangle \leftrightarrow\left\vert
2\right\rangle $ with rate $\Gamma_{ad}$. For an atomic vapor in the presence
of a buffer gas, $\Gamma_{n}$ will be the pressure-broadened homogenous width.
The equations of motion for the density matrix are thus%
\begin{align}
\dot{\rho}_{1,1} &  =-2\operatorname{Im}\Omega_{1}^{\ast}\left(  t\right)
\rho_{3,1}+\Gamma_{1}\rho_{3,3}+\Gamma_{2\leftrightarrow1}\left(  \rho
_{2,2}-\rho_{1,1}\right)  \nonumber\\
\dot{\rho}_{2,2} &  =-2\operatorname{Im}\Omega_{2}^{\ast}\left(  t\right)
\rho_{3,2}+\Gamma_{2}\rho_{3,3}-\Gamma_{2\leftrightarrow1}\left(  \rho
_{2,2}-\rho_{1,1}\right)  \nonumber\\
\dot{\rho}_{3,3} &  =2\operatorname{Im}\left[  \Omega_{1}^{\ast}\left(
t\right)  \rho_{3,1}+\Omega_{2}^{\ast}\left(  t\right)  \rho_{3,2}\right]
-\left(  \Gamma_{1}+\Gamma_{2}\right)  \rho_{3,3}\nonumber\\
\dot{\rho}_{2,1} &  =i\Omega\,_{2}^{\ast}\left(  t\right)  \rho_{3,1}%
-i\Omega\,_{1}\left(  t\right)  \rho_{2,3}-i\left(  \omega_{21}-i\Gamma
_{21}\right)  \rho_{2,1}\nonumber\\
\dot{\rho}_{3,1} &  =-i\Omega_{1}\left(  t\right)  \left(  \rho_{3,3}%
-\rho_{1,1}\right)  +i\Omega_{2}\left(  t\right)  \rho_{2,1}\nonumber\\
&  -i\left(  \omega_{31}-i\Gamma_{C}\right)  \rho_{3,1}\nonumber\\
\dot{\rho}_{3,2} &  =-i\Omega_{2}\left(  t\right)  \left(  \rho_{3,3}%
-\rho_{2,2}\right)  +i\Omega_{1}\left(  t\right)  \rho_{1,2}\nonumber\\
&  -i\left(  \omega_{32}-i\Gamma_{C}\right)  \rho_{3,2}\label{eom}%
\end{align}
where $\Gamma_{C}=\left(  \Gamma_{1}+\Gamma_{2}+\Gamma_{2\leftrightarrow
1}+\Gamma_{ad}\right)  /2$ , $\Gamma_{21}=\Gamma_{2\leftrightarrow1}%
+2\Gamma_{ad}$. To find the absorption spectrum of the probe
field, we follow Eqs.(\ref{2l_wdot})-(\ref{2l_s}) using the
response terms of the probe and get,
\begin{equation}
S\left(  \omega_{1}\right)  =\operatorname{Im}\lim_{T\rightarrow\infty}%
\int_{0}^{T}\frac{dt}{T}\frac{\rho_{3,1}\left(  t\right)  }{\Omega_{1}\left(
t\right)  }.\label{S_by_rho}%
\end{equation}

We consider the standard weak probe case, namely
$\Omega_{1}\ll\Omega_{2},$ and avoid saturation by assuming
$\Omega_{2}$ smaller than any of the relaxation rates
$\Gamma_{n\rightarrow n^{\prime}}.$ The perturbation solution of
Eqs.(\ref{eom}) can easily be done, following the same procedure
as in the two-level case. To zero order in $\Omega_{1},$ with the
initial state $\rho\left(  0\right) =\left\vert 1\right\rangle
\left\langle 1\right\vert ,$ we find $\rho_{1,1}^{(0)}=1$ and all
other matrix elements vanish. To the
first order in $\Omega_{1}$,%
\begin{subequations}
\begin{align}
\dot{\rho}_{3,1}^{(1)} &  =i\Omega_{2}\left(  t\right)  \rho_{2,1}%
^{(1)}+i\Omega_{1}\left(  t\right)  -i\left(  \omega_{31}-i\Gamma_{1}\right)
\rho_{3,1}^{(1)}\label{rho31_dot}\\
\dot{\rho}_{2,1}^{(1)} &  =i\Omega\,_{2}^{\ast}\left(  t\right)  \rho
_{3,1}^{(1)}-i\left(  \omega_{21}-i\Gamma_{21}\right)  \rho_{2,1}%
^{(1)}\label{rho21_dot}%
\end{align}
and all other matrix elements vanish. We solve formally Eq.(\ref{rho21_dot}),
assuming the initial state vanishes,%
\end{subequations}
\begin{equation}
\rho_{2,1}^{(1)}\left(  t\right)  =i\int_{0}^{t}dt_{1}\Omega_{2}^{\ast}\left(
t_{1}\right)  e^{\left(  -i\omega_{21}-\Gamma_{21}\right)  \left(
t-t_{1}\right)  }\rho_{3,1}^{(1)}\left(  t_{1}\right)  ,
\end{equation}
and substitute it back into Eq.(\ref{rho31_dot}),%
\begin{align}
\dot{\rho}_{3,1}^{(1)} &  =i\Omega_{1}\left(  t\right)  -i\left(  \omega
_{31}-i\Gamma_{1}\right)  \rho_{3,1}^{(1)}\\
&  -\Omega_{2}\left(  t\right)  \int_{0}^{t}dt_{1}\Omega_{2}^{\ast}\left(
t_{1}\right)  e^{\left(  -i\omega_{21}-\Gamma_{21}\right)  \left(
t-t_{1}\right)  }\rho_{3,1}^{(1)}\left(  t_{1}\right)  ,\nonumber
\end{align}
whose formal solution is an integral equation for $\rho_{3,1}^{(1)}\left(
t\right)  ,$
\begin{align}
\rho_{3,1}^{(1)}\left(  t\right)   &  =\int_{0}^{t}dt_{2}e^{\left(
-i\omega_{31}-\Gamma_{1}\right)  \left(  t-t_{2}\right)  }\left[  i\Omega
_{1}\left(  t_{2}\right)  -\right.  \label{rho31_1}\\
&  \left.  \Omega_{2}\left(  t_{2}\right)  \int_{0}^{t_{2}}dt_{1}\Omega
_{2}^{\ast}\left(  t_{1}\right)  e^{\left(  -i\omega_{21}-\Gamma_{21}\right)
\left(  t_{2}-t_{1}\right)  }\rho_{3,1}^{(1)}\left(  t_{1}\right)  \right]
.\nonumber
\end{align}
An approximate solution for Eq.(\ref{rho31_1}) is obtained by iterations up to
first order in $\rho_{3,1}$. It can be shown, by comparing the approximate
solution to the exact solution for an atom at rest, that the approximation is
valid in the \emph{low-contrast} regime, when $\left\vert \Omega
_{2}\right\vert ^{2}\ll\Gamma_{12}\Gamma$, i.e. in the limit of small relative
increase in transparency. Note that the power-broadening in this regime is
small compared to $\Gamma_{12}$.

Performing two iterations of Eq.(\ref{rho31_1}) and returning to
Eq.(\ref{S_by_rho}), we find the spectrum to be the sum of two terms: (i) The
one-photon absorption spectrum,%
\begin{equation}
S_{1}=\int_{0}^{\infty}d\tau e^{-\Gamma_{1}\tau-\Gamma_{D}^{2}G\left(
\gamma\tau\right)  /\gamma^{2}}\cos\left(  \Delta_{1}\tau\right)
\end{equation}
where $\Gamma_{D}=q_{1}v_{\text{th}}$ and $\Delta_{1}=\omega_{31}-\omega_{1}$
is the one-photon detuning of the probe transition. This is essentially the
two-level absorption spectrum, Eq.(\ref{2l_s_final}), and is equivalent to the
small-signal absorption spectrum in the absence of the pump. (ii) The
two-photon absorption dip (the CPT line-shape)%
\begin{align}
S_{2}  &  =-\left\vert \Omega_{2}\right\vert ^{2}\operatorname*{Re}%
\lim_{T\rightarrow\infty}\int_{0}^{T}\frac{dt}{T}\int_{0}^{t}dt_{1}\int
_{0}^{t_{1}}dt_{2}\int_{0}^{t_{2}}dt_{3}\nonumber\\
&  e^{\left(  -i\Delta_{1}-\Gamma_{1}\right)  \left(  t-t_{1}+t_{2}%
-t_{3}\right)  }e^{\left(  i\Delta_{R}-\Gamma_{21}\right)  \left(  t_{1}%
-t_{2}\right)  }\times\nonumber\\
&  e^{i\mathbf{q}_{2}\mathbf{\cdot}\left[  \mathbf{r}\left(  t_{1}\right)
-\mathbf{r}\left(  t_{2}\right)  \right]  -i\mathbf{q}_{1}\mathbf{\cdot
}\left[  \mathbf{r}\left(  t\right)  -\mathbf{r}\left(  t_{3}\right)  \right]
}%
\end{align}
where $\Delta_{R}=\omega_{1}-\omega_{21}-\omega_{2}$ is the Raman detuning.
For brevity, we will take $\Delta_{1}=0$ and write the spectrum in terms of
$\Delta_{R}$, i.e. assuming the CPT resonance to occur around the center of
the one-photon absorption line. The calculation can easily be generalized to
$\Delta_{1}\neq0$.

To calculate the shape of the absorption dip, we write it as%
\begin{align}
S_{2}\left(  \Delta_{R}\right)   &  =-\left\vert \Omega_{2}\right\vert
^{2}\operatorname*{Re}\lim_{T\rightarrow\infty}\int_{0}^{T}\frac{dt}{T}%
\int_{0}^{t}d\tau e^{\left(  i\Delta_{R}-\Gamma_{21}\right)  \tau}\nonumber\\
&  \int_{0}^{t-\tau}d\tau_{1}e^{-\Gamma_{1}\tau_{1}}\int_{0}^{\tau_{1}}%
d\tau_{3}e^{iK}%
\end{align}
where
\[
K\equiv\Phi_{2}\left(  t-\tau_{3},\tau\right)  -\Phi_{1}\left(  t,\tau
_{1}+\tau\right)
\]
and $\Phi_{n}\left(  t,\tau\right)  \equiv\mathbf{q}_{n}\mathbf{\cdot}\left[
\mathbf{r}\left(  t\right)  -\mathbf{r}\left(  t-\tau\right)  \right]  .$ Both
the integrations over $\tau$ and $\tau_{1}$ contain exponential decay terms,
allowing us to take their upper limits to infinity and average the phase-lag
term over $t$,%
\begin{align}
S_{2}\left(  \Delta_{R}\right)   &  =-\left\vert \Omega_{2}\right\vert
^{2}\operatorname*{Re}\int_{0}^{\infty}d\tau e^{\left(  i\Delta_{R}%
-\Gamma_{21}\right)  \tau}\nonumber\\
&  \int_{0}^{\infty}d\tau_{1}e^{-\Gamma_{1}\tau_{1}}\int_{0}^{\tau_{1}}%
d\tau_{3}\overline{e^{iK}}.\label{S1_1}%
\end{align}
In a similar manner to the two-level analysis, we use the cumulant expansion
to get%
\begin{equation}
\overline{e^{iK}}\approx e^{-\overline{K^{2}}/2}.\label{eik}%
\end{equation}
Then%
\begin{equation}
\overline{K^{2}}=\overline{\Phi_{2}^{2}\left(  0,\tau\right)  }+\overline
{\Phi_{1}^{2}\left(  0,\tau_{1}+\tau\right)  }-2\overline{\Phi_{2}\Phi_{1}%
}\label{k_sqr}%
\end{equation}
where the first two terms can be found from Eq.(\ref{2l_phisqr}) and the third
term is%
\begin{equation}
\overline{\Phi_{2}\Phi_{1}}=\overline{\Phi_{2}\left(  -\tau_{3},\tau\right)
\Phi_{1}\left(  0,\tau_{1}+\tau\right)  }.
\end{equation}
\bigskip Using Eqs.(\ref{2l_single_phi}),(\ref{2l_uu_corr}) we find%
\begin{align}
\overline{\Phi_{2}\Phi_{1}}  & =\sum_{\alpha,\alpha^{\prime}}q_{1}^{\alpha
}q_{2}^{\alpha^{\prime}}\int_{-\tau_{1}-\tau}^{0}dt_{1}\int_{-\tau_{3}-\tau
}^{-\tau_{3}}dt_{2}\overline{u^{\alpha}\left(  t_{1}\right)  u^{\alpha
^{\prime}}\left(  t_{2}\right)  }\nonumber\\
& =\mathbf{q}_{1}\mathbf{\cdot q}_{2}v_{\text{th}}^{2}\int_{-\tau_{1}-\tau
}^{0}dt_{1}\int_{-\tau_{3}-\tau}^{-\tau_{3}}dt_{2}e^{-\gamma\left\vert
t_{2}-t_{1}\right\vert }.
\end{align}
Performing the integral and substituting the results back into Eq.(\ref{S1_1}%
), we get%
\begin{align}
S_{2}\left(  \Delta_{R}\right)   &  =-\left\vert \Omega_{2}\right\vert
^{2}\operatorname*{Re}\int_{0}^{\infty}d\tau e^{\left(  i\Delta_{R}%
-\Gamma_{21}\right)  \tau}e^{-\left(  \Gamma_{D}^{\text{res}}/\gamma\right)
^{2}G\left(  \gamma\tau\right)  }\nonumber\\
&  \int_{0}^{\infty}d\tau_{1}e^{-\Gamma_{1}\tau_{1}}e^{-\left(  \Gamma
_{D}/\gamma\right)  ^{2}\left(  \gamma\tau_{1}-e^{-\gamma\tau}+e^{-\gamma\tau
}e^{-\gamma\tau_{1}}\right)  }\nonumber\\
&  \int_{0}^{\tau_{1}}d\tau_{3}e^{\mathbf{q}_{1}\mathbf{\cdot q}%
_{2}v_{\text{th}}^{2}\left(  e^{-\gamma\tau}-1\right)  \left(  e^{-\gamma
\tau_{1}}e^{\gamma\tau_{3}}+e^{-\gamma\tau_{3}}-2\right)  /\gamma^{2}%
}\label{S2_0}%
\end{align}
where $G(x)$ was defined in Eq.(\ref{2l_Gx}) and%
\begin{equation}
\Gamma_{D}^{\text{res}}=\left\vert \mathbf{q}_{1}-\mathbf{q}_{2}\right\vert
v_{\text{th}}%
\end{equation}
is the residual Doppler width.

Eq.(\ref{S2_0}) is a general analytic expression for the CPT line-shape. In
what follows we limit the discussion to a specific realistic regime, which is
relevent to most CPT experiments. In realizations of CPT in vapor medium, the
one-photon processes are usually in the far Doppler regime ($\Gamma_{D}%
\gg\Gamma_{1}~$and $\Gamma_{D}\gg\gamma$) and in most applications the vapor
cell contains buffer gas, forcing the two-photon processes into the Dicke
regime ($\gamma\gg\Gamma_{D}^{\text{res}}\gg\Gamma_{12}$). Since this regime
includes both Doppler and Dicke regimes we denote it as the\emph{ intermediate
regime}. As described in the previous section, for the Doppler regime we take
$e^{-\gamma\tau_{1}}\approx1-\gamma\tau_{1},$ and thus $e^{-\gamma\tau_{3}%
}\approx1-\gamma\tau_{3}$ and for the Dicke regime we take $e^{-\gamma\tau}%
\ll1$ and $G\left(  \gamma\tau\right)  \approx\gamma\tau$. With these
approximations, a simple expression for the CPT line-shape is obtained%
\begin{equation}
S_{2}\left(  \Delta_{R}\right)  =\frac{-\left\vert \Omega_{2}\right\vert ^{2}%
}{\left[  \Gamma_{1}+\mathbf{q}_{1}\mathbf{\cdot}\left(  \mathbf{q}%
_{1}-\mathbf{q}_{2}\right)  v_{\text{th}}^{2}/\gamma\right]  ^{2}}\frac
{\Gamma_{12}+\eta\Gamma_{D}^{\text{res}}}{\Delta_{R}^{2}+\left[  \Gamma
_{12}+\eta\Gamma_{D}^{\text{res}}\right]  ^{2}}, \label{S2}%
\end{equation}
where the parameter%
\begin{equation}
\eta=\frac{\Gamma_{D}^{\text{res}}}{\gamma}=2\pi\frac{\Lambda}{\lambda
_{\text{CPT}}}%
\end{equation}
is proportional to the ratio between the mean free path, $\Lambda,$ and the
wavelength associated with the \emph{wave-vector difference}, $\lambda
_{\text{CPT}}=2\pi/\left\vert \mathbf{q}_{1}-\mathbf{q}_{2}\right\vert $.
Equation (\ref{S2}) is the main result of the present work, showing that the
CPT line-shape is the product of two terms, which are both functions of the
wave-vector difference. The first term determines the line's amplitude and the
second is a Lorentzian that determines the width. Since $\eta$ multiplies the
residual Doppler width, it acts as a narrowing factor and we denote it as the
CPT-Dicke parameter. Finally, in a typical setup of a CPT-based frequency
standard \cite{knappe:1460}, the laser beams are collinear $\left(
\mathbf{q}_{1}\parallel\mathbf{q}_{2}\right)  $ and the line-shape can be
written as%
\begin{equation}
S_{2}^{\parallel}\left(  \Delta_{R}\right)  =\frac{-\left\vert \Omega
_{2}\right\vert ^{2}}{\left(  \Gamma_{1}+\eta\Gamma_{D}\right)  ^{2}}%
\frac{\Gamma_{12}+\eta\Gamma_{D}^{\text{res}}}{\Delta_{R}^{2}+\left(
\Gamma_{12}+\eta\Gamma_{D}^{\text{res}}\right)  ^{2}}.
\end{equation}

\section{ Discussion and conclusions}

CPT\ is an inherently narrow-band phenomena usually limited by
effective broadening mechanisms. Non-degenerate CPT\ resonances in
a hot vapor cell with buffer gas would naively be broadened by a
residual Doppler broadening. However, the measured line-width of
CPT resonances are far below the expected residual Doppler width.
This effect was attributed to the frequent velocity changing
collisions with the buffer gas, that are well known to narrow
atomic absorption transitions in two-level atoms. It was also
demonstrated using a numerical simulation that when frequent
collisions occur no Doppler broadening is evident in the CPT line
\cite{HelmPRA2001}. In this work we developed the theory of Dicke
narrowing for CPT resonances in three level atoms. The main result
is that the residual Doppler width (that would be observable in an
apparatus with no collisions) is diminished by the ratio of the
mean free path between collisions and the wavelength associated
with the wave-vector difference of the two radiation fields. This
theory can be readily extended to describe atoms in confined
geometries such as thin vapor cells and cold atoms traps.

For hyperfine CPT experiments, performed in vapor cells with
Alkali atoms and several Torrs of buffer gas, the typical
CPT-Dicke parameter is $\eta \approx10^{-4}$ (i.e. very strong
Dicke narrowing). Hence the residual Doppler broadening, which is
of the order of a few KHz, is strongly reduced and is not
measurable (compared to other broadening mechanisms). In order to
verify the theoretical prediction it is necessary to increase
either the CPT-Dicke parameter or the residual Doppler width. An
increase of the CPT-Dicke parameter can be achieved by decreasing
the effective wavelength or by increasing the mean free path
between collisions. However, changing the effective wavelength is
limited by the atomic structure of the active atoms and changing
the mean free path will result in a significant
diffusion-broadening. We propose to increase the residual Doppler
width by introducing a small angular deviation, $\theta$, between
the pump and probe beams. For CPT performed with two degenerate
lower levels ($\left\vert \mathbf{q}_{1}\right\vert =\left\vert
\mathbf{q}_{2}\right\vert $) and small $\theta$, both the residual
Doppler width and the CPT-Dicke parameter are proportional to
$\theta$. Therefore, the resulting broadening is proportional to
$\theta^{2}$ and can be increased to a measurable level.

\begin{acknowledgments}
We thank Nitsan Aizenshtark for reading the manuscript and helpful
suggestions. This work was partially supported by DDRND\ and the fund for
encouragement of research in the Technion.
\end{acknowledgments}

\appendix

\section*{APPENDIX}

To obtain the velocity-velocity correlation function of\ Eq.(\ref{2l_phisqr0})
we review two simple models:

(i) In the \textit{Brownian motion} case the equation of motion of the
$\alpha$ component of the velocity is
\begin{equation}
\frac{d}{dt}u^{\alpha}\left(  t\right)  \mathbf{=-}\gamma u^{\alpha}\left(
t\right)  \mathbf{+}a^{\alpha}\left(  t\right)  \mathbf{,} \label{ut-1}%
\end{equation}
where $\gamma u^{\alpha}\left(  t\right)  $\ is the dynamic frictional force
on the atom, $\gamma$\ is the velocity relaxation rate, and $a^{\alpha}\left(
t\right)  $\ is the random acceleration \cite{Chandrasekhar1943}. For the
correlation function we thus obtain
\begin{equation}
\frac{d}{dt}\overline{u^{\alpha}\left(  t\right)  u^{\alpha^{\prime}}\left(
0\right)  }\mathbf{=-}\gamma\overline{u^{\alpha}\left(  t\right)
u^{\alpha^{\prime}}\left(  0\right)  }+\overline{a^{\alpha}\left(  t\right)
u^{\alpha^{\prime}}\left(  0\right)  }\mathbf{,} \label{ut-2}%
\end{equation}
where the bar indicates ensemble average. Since the acceleration is not
correlated with the initial velocity, the last term on the right vanishes, and
we have%
\begin{equation}
\overline{u^{\alpha}\left(  t\right)  u^{\alpha^{\prime}}\left(  0\right)
}=\overline{u^{\alpha}\left(  0\right)  u^{\alpha^{\prime}}\left(  0\right)
}e^{-\gamma\left\vert t\right\vert }. \label{ut-3}%
\end{equation}
In thermal equilibrium with temperature $T,$ the ensemble average is taken
over the velocity distribution function%
\begin{equation}
F(\mathbf{u})=\left(  m/2\pi k_{B}T\right)  ^{3/2}e^{-mu^{2}/2k_{B}T},
\label{bol-1}%
\end{equation}
where\ $m$\ is the mass of the atom, and $k_{B}$ is the Boltzmann factor,\ and
we get
\begin{equation}
\overline{u^{\alpha}\left(  0\right)  u^{\alpha^{\prime}}\left(  0\right)
}=\delta_{\alpha^{\prime}\alpha}v_{\text{th}}^{2}, \label{ut-4}%
\end{equation}
where $\delta_{\alpha,\alpha^{\prime}}$ is the Kronecker Delta, and
$v_{\text{th}}=\sqrt{k_{B}T/m}$ is the thermal velocity.

(ii) In the \textit{Strong Collisions }case the atom is colliding with the
dilute buffer gas in thermal equilibrium. The conditional probability density,
$f(\mathbf{u},t),$ to find the atom with velocity $\mathbf{u}$ at time $t,$
given that at time $t=0$ its velocity is $\mathbf{u}\left(  0\right)  ,$ is
given by the Boltzmann collision term,
\begin{equation}
\frac{d}{dt}f(\mathbf{u},t)=-\left[  \frac{d}{dt}f(\mathbf{u},t)\right]
_{coll}, \label{be-1}%
\end{equation}
with the initial distribution
\begin{equation}
f(\mathbf{u},0)=\delta\left(  \mathbf{u-u}\left(  0\right)  \right)  .
\label{fi-1}%
\end{equation}
The simplest model is the \textit{single relaxation rate} approximation,
where
\begin{equation}
\left[  \frac{d}{dt}f(\mathbf{u},t)\right]  _{coll}=\gamma\left(
f(\mathbf{u},t)-F(\mathbf{u})\right)  , \label{be-2}%
\end{equation}
$F(\mathbf{u})$ is the equilibrium distribution of Eq.(\ref{bol-1}), and
$\gamma$ is the collision relaxation rate. The solution of Eq.(\ref{be-1}),
with Eq.(\ref{be-2}), is simply
\begin{equation}
f(\mathbf{u},t)=F(\mathbf{u})\left(  1-e^{-\gamma\left\vert t\right\vert
}\right)  +e^{-\gamma\left\vert t\right\vert }f(\mathbf{u},0), \label{be-4}%
\end{equation}
and, with Eq.(\ref{fi-1}), the velocity of the atom at time $t,$\ is
\begin{equation}
\mathbf{u}\left(  t\right)  =\int d^{3}u\mathbf{u}f(\mathbf{u},t)=\mathbf{u}%
\left(  0\right)  e^{-\gamma\left\vert t\right\vert }. \label{vt-1}%
\end{equation}
The velocity-velocity correlation function is
\begin{equation}
\overline{u^{\alpha}\left(  t\right)  u^{\alpha^{\prime}}\left(  0\right)
}=\overline{u^{\alpha}\left(  0\right)  u^{\alpha^{\prime}}\left(  0\right)
}e^{-\gamma\left\vert t\right\vert }. \label{ut-5}%
\end{equation}
and since
\begin{equation}
f(\mathbf{u},t\rightarrow\infty)=F(\mathbf{u}), \label{fm-2}%
\end{equation}
for consistency
\begin{equation}
\overline{u^{\alpha}\left(  0\right)  u^{\alpha^{\prime}}\left(  0\right)
}=\int d^{3}uu^{\alpha}u^{\alpha^{\prime}}F(\mathbf{u})=\frac{k_{B}T}{m}.
\label{ut-6}%
\end{equation}
We end up in both these models with%
\begin{equation}
\overline{u^{\alpha}\left(  t\right)  u^{\alpha^{\prime}}\left(  0\right)
}=\delta_{\alpha^{\prime}\alpha}v_{\text{th}}^{2}e^{-\gamma\left\vert
t\right\vert }.
\end{equation}
where the interpretation of $\gamma$ is either the velocity relaxation rate or
the\ collision relaxation rate.

\bibliographystyle{apsrev}
\bibliography{references}

\end{document}